\documentclass[twocolumn,english,aps,prl,showpacs]{revtex4}
\usepackage[T1]{fontenc}
\usepackage[latin1]{inputenc}
\usepackage{amsmath}
\usepackage{amssymb}

\makeatletter
\usepackage[dvips]{graphicx}

\input{epsf}

\usepackage{babel}

\usepackage{babel}

\usepackage{babel}

\usepackage{babel}
\makeatother

\begin{document}

\title{Density fingerprint of giant vortices in Fermi gases near a Feshbach resonance}

\author{Hui Hu$^{1,2}$ and Xia-Ji Liu$^{2}$} 

\affiliation{$^{1}$\ Department of Physics, Renmin University of China, Beijing 100872, China \\
$^{2}$\ ARC Centre of Excellence for Quantum-Atom Optics, Department of Physics, 
University of Queensland, Brisbane, Queensland 4072, Australia}

\date{\today{}}

\begin{abstract}
The structure of multiply quantized or giant vortex states in atomic Fermi gases across 
a Feshbach resonance is studied within the context of self-consistent Bogoliubov-de 
Gennes theory. The particle density profile of vortices with $\kappa>1$ flux quanta is 
calculated. Owing to $\kappa$ discrete branches of vortex core bound states,  inside 
the core the density oscillates as a function of the distance from the vortex line and displays 
a non-monotoic dependence on the interaction strengths, in marked contrast to the singly 
quantized case, in which the density depletes monotonically. This feature, never reported 
so far, can make a direct visualization of the giant vortices in atomic Fermi gases.
\end{abstract}


\maketitle

One of the hallmarks of superfluidity of quantum fluids, be it fermionic or 
bosonic, is the appearance of quantized vortices. In condensed matter physics, 
the vortex states have been studied widely in various systems, 
ranging from the conventional Bardeen-Cooper-Schrieffer (BCS) superconductors 
to the rotating helium superfluids. The recent manipulated ultracold atomic $^6$Li 
and $^{40}$K gases emerge as a new promising test-bed for the vortex 
physics \cite{crossover}. Their interactions can be arbitrarily and precisely 
enhanced using a Feshbach resonance. Upon sweeping a magnetic field downward 
through the resonance, these Fermi systems undergo a smooth crossover from the 
BCS superfluidity to the Bose-Einstein condensation (BEC) of tightly bound pairs 
\cite{crossover}. As the underlying statistics of systems changes from fermionic 
to bosonic across the resonance, it is interesting to ask how the properties of 
vortices evolve around the crossover.

A singly quantized vortex in the BCS-BEC crossover has been discussed to a certain 
extent \cite{hayashi,bruun,nygaard,bulgac,tempere,machida,chien,sensarma}. The presence of strong interactions 
is shown to lead to a significant depletion of the particle density in the region 
of the vortex core \cite{hayashi,bulgac}, which was indeed confirmed experimentally by 
Zwierlein \textit{et al.} for a $^6$Li gas \cite{crossover}. The properties of 
giant vortex states with multiple flux quanta \cite{virtanen,tanaka,duncan}, on the 
other hand, is less known. Generically, in a bulk system giant vortices are not
energetically favorable and are not expected to persist if created. In the
confined geometry, however, the situation may be different. A number of methods 
have been proposed to overcome this vortex dissociation instability, including the 
use of an external repulsive pinning potential \cite{simula} or a trapping potential 
steeper than the harmonic traps \cite{fetter}. As a counterpart, the giant vortex 
structures have been recently produced in rapid rotating trapped BECs \cite{engels}. 
They have also been observed in the nanoscale superconductors where the sample 
size becomes comparable to the superconducting coherence length $\xi $ \cite{kanda}.

Given all the recent advances in experimental techniques, in this paper we discuss 
the evolution of the giant vortex structure from weak-to-strong coupling superfluidity 
in trapped Fermi gases across a broad Feshbach resonance. In marked contrast to 
the singly quantized vortex, we find a non-trivial oscillation behavior in the 
particle density profile of giant vortices inside the core in the strongly interacting 
BCS-BEC crossover regime. The oscillation pattern, unique to the number of flux
quanta at particular couplings, relates directly to the microscopic electronic spectrum 
of the local density of states (LDOS), which acquires an intriguing structure owing to 
the multiple branches of vortex core bound states, the so-called Caroli-de 
Gennes-Matricon (CdGM) states \cite{caroli}. In this respect, it provides a density 
\emph{fingerprint} for giant vortices in the neutral Fermi gases. Towards the deep 
BEC limit, these oscillation patterns cease to exist, and finally the density profile
returns back to that of a BEC. Our results are obtained by numerically solving the 
Bogoliubov-de Gennes (BdG) equations in a fully self-consistent fashion. As the strongly 
interacting Fermi systems can be found also in various fields of physics, such as the 
high-temperature superconductors and neutron stars, our results can have implications 
beyond the cold atom physics.

To be concrete, we consider a two-dimensional (2D) Fermi gas that can be prepared readily 
in a single `pancake' trap or at the nodes of 1D optical lattice potentials. 
It is sufficient to model the broad Feshbach resonance using a single channel Hamiltonian 
\cite{xiaji}. We therefore assume a 2D contact interaction parameterized by a coupling 
constant $g$. The two-body interaction problem under this circumstance involves two 
length scales: the characteristic length in the tightly confined direction $a_0$ and the 
3D $s$-wave scattering length $a_{sc}$. A peculiar 2D bound state of two atoms appears for 
an \emph{arbitrarily} weak attraction \cite{petrov}, with the binding energy 
$E_a/(\hbar \omega _0)=0.915/\pi \exp (-\sqrt{2\pi }a_0/a_{sc})\ll 1$, 
where $\omega _0=\hbar /(ma_0^2)$. The bare coupling constant can then be regularized via the 
$s$-wave scattering phase shift \cite{randeria}, \textit{i.e.}, 
\begin{equation}
\frac 1g+\sum_{\mathbf{k}}\frac 1{\hbar^2 \mathbf{k}^2/m + E}=\frac m{4\pi \hbar ^2}\ln \left( \frac{E_a}E\right) ,  \label{regularization}
\end{equation}
where the relative collision energy $E$ is of the order of the Fermi energy $E_F$ and drops 
automatically out of the final results. For a uniform gas at zero temperature, the 
mean-field theory of the BCS-BEC crossover in 2D admits simple analytic expressions for 
the order parameter and chemical potential: $\Delta =(2E_FE_a)^{1/2}$ and $\mu =E_F-E_a/2$, 
respectively \cite{randeria}. Hence, $E_a\ll E_F$ corresponds to the weak coupling BCS limit,
while in the opposite BEC limit of very strong attractions, $E_a\gg E_F$. The crossover 
occurs approximately at $E_a\simeq 0.5E_F$ \cite{footnote}.

In BdG approach the quasiparticle wave functions $u_\eta$ and $v_\eta$ are 
determined by the coupled equations \cite{gygi},
\begin{equation}
\left[ 
\begin{array}{cc}
{\cal H}_0 & \Delta (\mathbf{r}) \\ 
\Delta ^{*}(\mathbf{r}) & -{\cal H}_0
\end{array}
\right] \left[ 
\begin{array}{c}
u_\eta \left(\mathbf{r}\right) \\ 
v_\eta \left(\mathbf{r}\right)
\end{array}
\right] =E_\eta \left[
\begin{array}{c}
u_\eta \left(\mathbf{r}\right) \\ 
v_\eta \left(\mathbf{r}\right)
\end{array}
\right] ,  \label{BdG}
\end{equation}
where $E_\eta $ is the excitation energy, and the single particle
Hamiltonian in traps is ${\cal H}_0=-\hbar ^2 \mathbf{\nabla}^2/2m+m\omega ^2r^2/2-\mu $. 
As the BdG equations are invariant under the replacement $u_\eta \left(\mathbf{r}\right)
\rightarrow v_\eta ^{*}\left(\mathbf{r}\right) $, $v_\eta \left(\mathbf{r}\right) \rightarrow 
-u_\eta ^{*}\left(\mathbf{r}\right) $, $E_\eta \rightarrow -E_\eta $, we restrict our 
calculations to $E_\eta \geq 0$ only. The order parameter $\Delta (\mathbf{r})$ and the 
chemical potential $\mu $ are determined respectively by the self-consistency equation 
$\Delta (\mathbf{r})=g\sum_\eta u_\eta \left(\mathbf{r}\right) v_\eta ^{*}\left(\mathbf{r}\right)[1-2f(E_\eta )]$ 
and the particle density $n\left(\mathbf{r}\right) =2\sum_\eta \{\left| u_\eta \left(\mathbf{r}\right) \right| ^2f(E_\eta )
+\left| v_\eta \left(\mathbf{r}\right) \right| ^2[1-f(E_\eta )]\}$ so that $\int d\mathbf{r}n\left(\mathbf{r}\right) =N$. 
Here $f\left( x\right) =1/\left( e^{x/k_BT}+1\right) $ is the Fermi distribution function and $N$ is the number 
of total atoms. Numerically one has to truncate the summation over the energy levels. 
In practice, we develop a \emph{hybrid} procedure by introducing a high energy 
cut-off $E_c$, above which we use a local density approximation (LDA) for the 
high-lying modes. Thus the regularization prescription (\ref{regularization}) leads 
naturally to an effective coupling constant in the self-consistency equation 
$\Delta (\mathbf{r})=g_{eff}\left(\mathbf{r}\right) \sum_\eta u_\eta \left(\mathbf{r}\right)
v_\eta ^{*}\left(\mathbf{r}\right) [1-2f(E_\eta )]$, where $\sum_\eta $ is now
restricted to $E_\eta \leq E_c$. Further expression of $g_{eff}\left(\mathbf{r}\right) $ 
and the detailed LDA contributions to the particle density will be reported elsewhere. 
Below $E_c$, we solve the BdG equations by taking $\Delta (\mathbf{r})=\Delta (r)e^{-i\kappa \varphi }$, 
where $\kappa $ denotes the number of vortex flux quanta. Accordingly, we write, 
for the normalized modes, $u_\eta \left(\mathbf{r}\right) =u_{nm}\left( r\right) e^{i(m)\varphi }/\sqrt{2\pi }$ 
and $v_\eta \left(\mathbf{r}\right) =v_{nm}\left(r\right) e^{i\left( m+\kappa\right) \varphi }/\sqrt{2\pi }$. 
The BdG equations then decouple into different $m$ sectors and reduce to a matrix 
diagonalization problem if one expands $u_{nm}\left( r\right) $ and $v_{nm}\left( r\right) $ 
in a basis set of 2D harmonic oscillators.

We have performed self-consistent computations for a gas with $N=1000$ for $\kappa $ 
up to 10, and have set $a_{ho}=(\hbar /m\omega )^{1/2}$ and $\hbar\omega $ as the units 
of length and energy, respectively. In the absence of vortices, an interesting aspect 
of the 2D mean-field theory is that the density profile is essentially \emph{independent} 
on the interactions, though the chemical potential is appreciably reduced. Within 
LDA we find that $n\left(\mathbf{r}\right) _{\kappa =0}=(\sqrt{N}/\pi )(1-r^2/r_{TF}^2) a^{-2}_{ho}$
with $r_{TF}=\sqrt{2}N^{1/4} a_{ho}$, and $\mu _{\kappa =0}=E_F-E_a/2$, where $E_F=\sqrt{N} \hbar \omega \equiv k_BT_F$. 
The resulting maximum value of the order parameter is $\Delta _0=\left( 2E_a/E_F\right) ^{1/2}E_F$. 
We have chosen $E_c\simeq4E_F$, which is already sufficient large to ensure the cut-off 
independence of our results. The characteristic length scale of the core size of 
giant vortices is $\xi _\kappa \simeq $ $\kappa \xi $, where the coherence length 
$\xi \simeq \hbar v_F/\pi \Delta _0\sim k_F^{-1}=\sqrt{1/2}N^{-1/4}a_{ho}$ in the BCS-BEC 
crossover regime.

%%%%%%%%%%%%%%%%%%%%%%%%%%%%%%%%%%%%%%%%%%%%%%%%%%%%%%%%%%%%%%%%%%%%%%%%%%%%%%%%%
\begin{figure}[tbp]
\begin{center}\includegraphics*[width=0.45\textwidth]{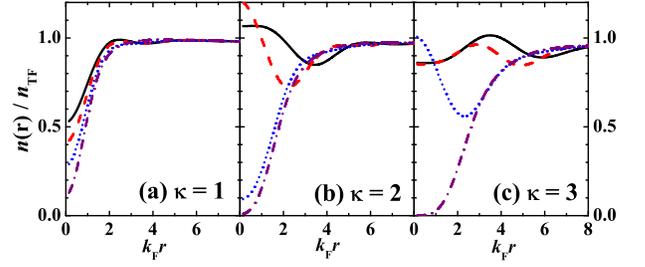}\end{center}

\caption{(Color online). Particle density profiles, normalized 
by $n_{TF}=\sqrt{N}/\pi a^{-2}_{ho}$,  for several values of the interaction strengths: 
$E_a=0.1E_F$ (solid lines), $E_a=0.2E_F$ (dashed lines), $E_a=0.5E_F$ (dotted
lines), and $E_a=2.0E_F$ (dash-dotted lines).}

\label{fig1}
\end{figure}
%%%%%%%%%%%%%%%%%%%%%%%%%%%%%%%%%%%%%%%%%%%%%%%%%%%%%%%%%%%%%%%%%%%%%%%%%%%%%%%%%

%%%%%%%%%%%%%%%%%%%%%%%%%%%%%%%%%%%%%%%%%%%%%%%%%%%%%%%%%%%%%%%%%%%%%%%%%%%%%%%%%
\begin{figure}[tbp]
\begin{center}\includegraphics*[width=0.45\textwidth]{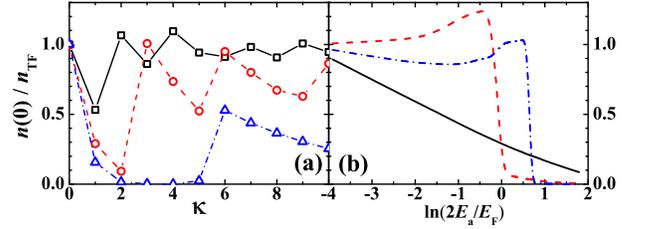}\end{center}

\caption{(Color online). Centre particle density as a function of the number of flux quanta (a)
and the interaction strengths (b). The symbols in the left panel denote different value of the 
interaction strengths: $E_a=0.1E_F$ (BCS side, squares), $E_a=0.5E_F$ (crossover point, circles), and 
$E_a=1.5E_F$ (BEC side, triangles). The lines in the right panel show the results with different 
number of flux quanta: $\kappa =1$ (solid line), $\kappa =2$ (dashed line), 
and $\kappa =3$ (dash-dotted line).}

\label{fig2}
\end{figure}
%%%%%%%%%%%%%%%%%%%%%%%%%%%%%%%%%%%%%%%%%%%%%%%%%%%%%%%%%%%%%%%%%%%%%%%%%%%%%%%%%

Our main results are summarized in Figs. 1 and 2 where we report the vortex
particle density profiles and centre particle densities for several values of the number 
of flux quanta and the interaction strengths at nearly zero temperature $T=0.01T_F$. 
The most unexpected feature of these profiles is the prominent oscillation
behavior in the region of the vortex axis for giant vortices with $\kappa\geq 2$, 
in sharp contrast to the monotonic depletion of the particle density in case of 
a singly quantized vortex as shown in Fig. 1a. In addition, the centre particle density oscillates 
with $\kappa $ and displays a non-monotonic dependence on the interactions. 
The oscillations in the density profile are most pronounced on the BCS side and at large number of 
flux quanta. However, they get suppressed appreciably with increasing the interaction strengths. 
Nevertheless, they are clearly visible around the crossover regime, and should be 
easily detected by the absorption imaging in experiments. For a given interaction strength,
there is a critical value of the number of flux quanta required to exhibit the oscillations,
which increases as the interaction increases. In the nearly BEC regime at $E_a=1.5E_F$, the 
oscillation occurs for $\kappa \geq 6$ only, as displayed by the triangles in Fig. 2a. 
We thus expect that in the extreme BEC limit, these oscillation patterns in the density 
profiles of all giant vortices should vanish identically, in accordance with the general 
picture of a fully condensed BEC.

The appearance of the intriguing oscillations in the particle density profile 
for giant vortices is in close connection to the multiple branches of CdGM bound 
states inside the vortex core \cite{caroli}. In the weak coupling limit, a simple 
semiclassical treatment of the CdGM states leads to a linear spectrum \cite{duncan}, 
\begin{equation}
\epsilon _{nm}=(n+\frac 12 - \frac \kappa 2)E_{\kappa 0}+(m+ \frac \kappa 2)\kappa E_{\kappa 1},  \label{spectrum}
\end{equation}
where $m$ is the angular momentum, and the branch index $n$ may take $\kappa$ 
integrate values, \textit{i.e.}, from $-\kappa /2$ to ($\kappa -1)/2,$
according to the index theorem established by Volovik for the number of
anomalous branches of low-energy quasiparticles inside the core \cite{volovik}. 
$E_{\kappa 0}=\pi \hbar v_F/(2\xi _\kappa )\sim 2\Delta _0/\kappa$ and 
$E_{\kappa 1}=\hbar ^2/(2m\xi _\kappa ^2)\sim (\Delta _0^2/E_F)/\kappa ^2$ 
are the bound state level spacings \cite{duncan}.

%%%%%%%%%%%%%%%%%%%%%%%%%%%%%%%%%%%%%%%%%%%%%%%%%%%%%%%%%%%%%%%%%%%%%%%%%%%%%%%%%
\begin{figure}[tbp]
\begin{center}\includegraphics*[width=0.30\textwidth]{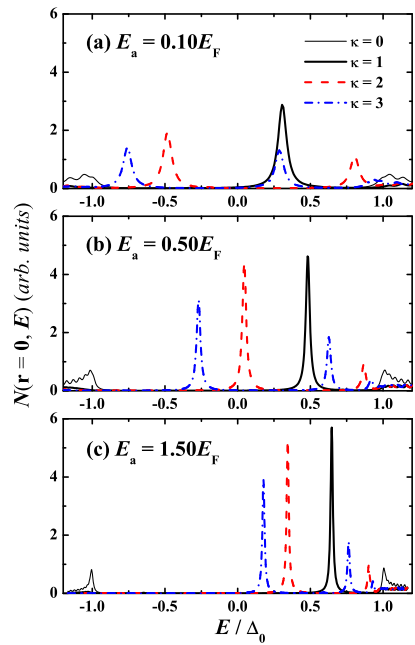}\end{center}

\caption{(Color online). Local fermionic density of states at the vortex axis 
for (a) $E_a=0.1E_F$, (b) $E_a=0.5E_F$, (c) and $E_a=1.5E_F$.}

\label{fig3}
\end{figure}
%%%%%%%%%%%%%%%%%%%%%%%%%%%%%%%%%%%%%%%%%%%%%%%%%%%%%%%%%%%%%%%%%%%%%%%%%%%%%%%%%

To illustrate the relation between CdGM states and our results on the particle 
density profiles, we calculate the LDOS $N\left(\mathbf{r},E\right) $, given by 
$2\sum_\eta [\left| u_\eta (\mathbf{r})\right| ^2\delta (E-E_\eta )+\left| v_\eta (\mathbf{r})\right| ^2\delta (E+E_\eta )]$, 
which, when integrated over negative energy, gives rise to the density profiles 
$n\left(\mathbf{r}\right)$. The CdGM states would exhibit themselves as peaks in 
the LDOS. As the radial functions behave as 
$u_{nm}\left( r\right) \sim r^{\left| m+\kappa /2\right| }$ and 
$v_{nm}\left( r\right) \sim r^{\left| m-\kappa /2\right| }$ close to the origin, 
the quasiparticle probability amplitudes $\left| u\left(\mathbf{r}\right) \right|^2$ and 
$\left| v\left(\mathbf{r}\right) \right| ^2$ have maxima at $r\simeq\left| m\right| /k_F$ 
because of the angular momentum of the states \cite{virtanen}. Therefore, the 
principal contribution to the LDOS at given $(\mathbf{r},E)$ arises from the bound 
states with $(\left| m\right| ,\epsilon_{nm})=\left( k_Fr,E\right) $ \cite{virtanen}.

Let us first focus on the centre particle density with $m\sim 0$. Fig. 3
shows the LDOS at the vortex axis for different values of the interactions.
On the BCS side, where $E_{\kappa 0}\gg E_{\kappa 1}$ (see, \textit{i.e.}, Fig.
3a), there are peaks located both below and above the Fermi surface of $E=0$
for $\kappa \geq 2$, and their weights may change periodically as a function
of $\kappa $. As a result, the centre density oscillates with the number of
flux quanta. With increasing the interactions, however, the level spacing 
$E_{\kappa 1}$ becomes progressively larger due to the enhancement of $\Delta_0$, 
and therefore $E_{\kappa 0}<E_{\kappa 1}$ across the crossover point. Hence, the 
interaction turns to expel the bound states towards the positive energy side. 
This results a sudden drop of the centre density at a critical coupling strength 
once the lowest bound state shifts up across $E=0$, as shown in Fig. 2b.
The value of the critical coupling increases with the number of flux quanta.

%%%%%%%%%%%%%%%%%%%%%%%%%%%%%%%%%%%%%%%%%%%%%%%%%%%%%%%%%%%%%%%%%%%%%%%%%%%%%%%%%
\begin{figure}[tbp]
\begin{center}\includegraphics*[width=0.40\textwidth]{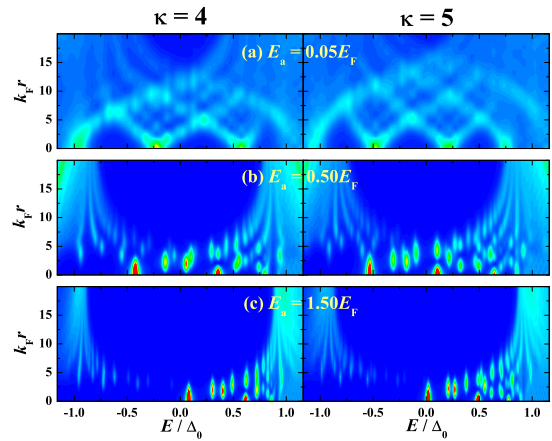}\end{center}

\caption{(Color online). Spatial variations of the local density of states for 
$\kappa=4$ and $\kappa =5$ at several values of the interaction strengths as
labeled.}

\label{fig4}
\end{figure}
%%%%%%%%%%%%%%%%%%%%%%%%%%%%%%%%%%%%%%%%%%%%%%%%%%%%%%%%%%%%%%%%%%%%%%%%%%%%%%%%%

We now consider the oscillations in the particle density profiles of giant
vortices, which may be understood from the spatial dependence of the LDOS,
as displayed in Fig. 4 for $\kappa =4$ and $\kappa =5$. In the weak coupling
limit, the $\kappa $ branch spectra of CdGM states are quasi-continuous. It
is easy to see from Eq. (\ref{spectrum}) that a wedge-shaped pattern of
maxima in the LDOS is formed \cite{virtanen,tanaka}, as shown in Fig. 4a.
There are $\kappa $ rows of peaks as one moves away from the vortex axis,
with decreasing number of peaks one by one due to the increase of the
angular momentum $m$. Therefore, the integration over the negative energy of
the LDOS naturally yields the oscillation behavior of the density profiles.
However, as noted above, the increase of the interaction will make the CdGM
states more discrete, with a larger level spacing. This destroys gradually
the regular pattern of maxima in the LDOS and the resulting oscillations in
the density profile. For a sufficient attraction, see, \textit{i.e.}, 
Fig. 4c, the LDOS exhausts at the negative energy, and therefore the density 
profile of giant vortices depletes completely inside the core, resembling that 
of an ideal BEC as expected.

%%%%%%%%%%%%%%%%%%%%%%%%%%%%%%%%%%%%%%%%%%%%%%%%%%%%%%%%%%%%%%%%%%%%%%%%%%%%%%%%%
\begin{figure}[tbp]
\begin{center}\includegraphics*[width=0.45\textwidth]{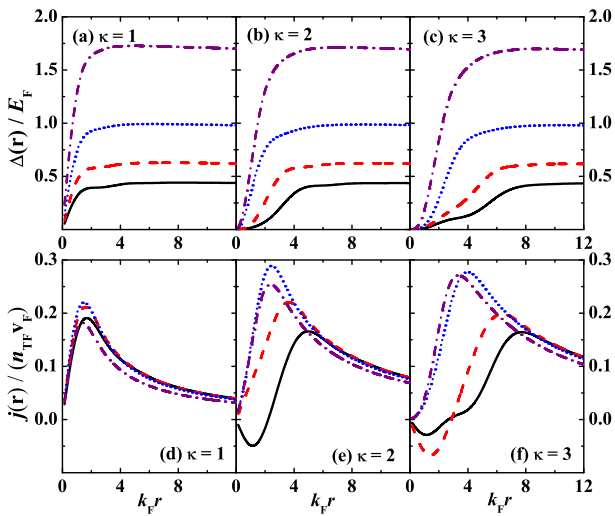}\end{center}

\caption{(Color online). Order parameter profiles and current distributions at the 
vortex core for several values of the interaction strengths: 
$E_a=0.1E_F$ (solid lines), $E_a=0.2E_F$ (dashed lines), $E_a=0.5E_F$ (dotted
lines), and $E_a=1.5E_F$ (dash-dotted lines). The currents are in units of
$n_{TF}v_F$, where $v_F=\hbar k_F /m$ is the Fermi velocity.}

\label{fig5}
\end{figure}
%%%%%%%%%%%%%%%%%%%%%%%%%%%%%%%%%%%%%%%%%%%%%%%%%%%%%%%%%%%%%%%%%%%%%%%%%%%%%%%%%

Finally, in Fig. 5 we report the order parameter profiles and the current 
circulating around the vortex core. Formally the current density is given by 
$\mathbf{j}\left(\mathbf{r}\right)=(2i\hbar/mr)\sum_{\eta}v_{\eta}\partial_{\varphi}v_{\eta}^{*}f(-E_{\eta})\mathbf{\hat{\varphi}}$.
The order parameter inside the core expands as the flux quanta increases, in accordance with the 
asymptotic Ginzburg-Landau form $\Delta(r) \sim r^{\kappa}$ ($r \lesssim \xi_\kappa$), and 
enhances with increasing the strength interactions. On the other hand, the current density 
exhibits a similar oscillation behavior as the particle density for weak interactions. 
These oscillations are attributed to the interplay between the paramagnetic bound states 
and the diamagnetic scattering states, which give the opposite contributions to the current,
as discussed in Ref. \cite{rainer} for a two-quantum vortex.

We so far confine to the 2D geometry. By allowing a free motion of atoms in a 
box of length $L$ in $z$ axis, we have also studied the 3D situation for 
a strongly interacting gas of $N=10^4$ atoms in a cylinder with $L \sim /r_{TF}$, 
and have observed qualitatively the same features.

In conclusion, by self-consistently solving the mean-field Bogoliubov-de Gennes equations 
we have analyzed the structure of giant vortices in a superfluid atomic Fermi gas in the 
strongly interacting BCS-BEC crossover regime. The multiple branches of the CdGM bound 
states are shown to have a significant impact on the local density of states, 
and consequently lead to nontrivial oscillations in the giant vortex density profiles. 
These distinct oscillations, which could be visualized after expanding the cloud, 
can make a useful diagnosis of giant vortices in atomic Fermi gases.

We acknowledge fruitful discussions with Professor Peter D. Drummond. This 
work was financially supported by the Australian Research Council Center of 
Excellence and by the National Science Foundation of China under Grant 
No. NSFC-10574080 and the National Fundamental Research Program under Grant
No. 2006CB921404.


\begin{thebibliography}{10}
\bibitem{footnote}In 3D the maximum value of the circulating current of
the singly quantized vortices, $j_{max}$, shows a non-monotonic behavior as a
function of $1/(k_Fa_{sc})$, peaking at unitarity \cite{sensarma}. Here we find the same
dependence of $j_{max}$ on $E_a/E_F$. Thus we locate the crossover point as
the peak position of $j_{max}$ and obtain $E_a\simeq 0.5E_F$.
\end{thebibliography}
\end{document}